\documentstyle[12pt,epsf]{article}
\newcommand{\text}[1]{\hbox{\rm \ #1\ \/}}

\newcommand{\RR}{\mbox{${\rm \:  R\!\!\!\! I
\;\;}$}} 

\newcommand{\qed}{\hfill $\Box$ \vskip 2ex}
\newcommand{\be}[1]{\begin{equation}\label{#1}}
\newcommand{\ee}{\end{equation}}
\newcommand{\vs}{\vspace{0.25cm}}

\begin{document}

\title{The Lie Algebra Rank Condition for 
Non-Bilinear Quantum Systems}

\author{Domenico D'Alessandro\\ 
Department of Mathematics\\
Iowa State University \\
Ames, IA 50011,  USA\\
Tel. (+1) 515 294 8130\\
email: daless@iastate.edu}

\maketitle

The controllability condition for right invariant systems on Lie
groups  derived in \cite{Suss} was  applied to quantum systems in 
\cite{confranot},  \cite{Tarn}, \cite{Rama}. This condition is   
called the {\it Lie Algebra Rank Condition}. In applications to quantum
systems, the condition  has been stated assuming
that the right invariant differential system under consideration 
is bilinear. The aim of this communication is to point out that 
this assumption is not necessary and in fact families 
of vector fields not necessarily bilinear have been already 
considered for
example in \cite{Jbook} (Chpt. 6). To make this communication
self-contained we give   
in the following a proof 
of this result. 

\vs 

We shall consider systems of the form 
\be{NOTBI}
\dot X=-iH(u_1,...,u_m)X, \qquad X(0)=I,   
\ee 
where   $H$ is a matrix (continuous) function of the 
controls $u_1,...,u_m$, which is assumed to be Hermitian 
for every value of $u_1,...,u_m$, in the control  set $U$,  
but otherwise arbitrary; The matrix $I$ is the $ n \times n$ Identity
matrix. We define the Lie Algebra $\cal L$ associated to system
(\ref{NOTBI}) as the Lie algebra generated by 
\be{generatori}
span_{ \{ u_1,...,u_m \} \in U} \{ iH(u_1,...,u_m) \}, 
\ee
and let $e^{\cal L}$ denote  the corresponding 
connected Lie group.  We shall denote
by $\cal R$ the set of states 
reachable from the Identity for system 
(\ref{NOTBI}). The admissible control functions are assumed to be
piecewise continuous functions with values in $U$. We have:

\vs

{ \bf Theorem:} 
\be{statement}
{\cal R}=e^{\cal L}.
\ee 

\vs

{\bf Proof.} ${\cal R} \subseteq e^{\cal L}$ follows from the fact
that $e^{\cal L}$ is the maximal integral manifold associated to the
vector fields defined in  (\ref{generatori}) [See Theorem 8.7 in
\cite{Boothby} and Lemma 2.4 in \cite{Suss}]. Therefore we have to prove
only that given an element $X_f$ of $e^{\cal L}$ there exist piecewise
constant controls such that the solution of (\ref{NOTBI}) attains the
value $X_f$. The topology of $e^{\cal L}$ is the one induced by the one
of $U(n)$. The proof uses arguments borrowed from 
\cite{miko}, \cite{sussJ}, \cite{Weaver}. We first state three  facts and
then conclude. 

\vs

{\it Fact 1:} { $\cal R$ is a semigroup}.

\vs

{\it Fact 2:} {$\cal R$ is dense in $e^{\cal L}$}

\vs

Let $A_1,...,A_s$ be a set of linearly independent 
generators of $\cal L$ obtained by setting
the controls equal to appropriate values in $iH(u_1,...,u_m)$ in
(\ref{generatori}). Clearly, 
all the elements in the one dimensional semigroups 
\be{semigruppi}
\{ X \in e^{\cal L} |X=e^{A_j t}, t \geq 0 \}, \quad j=1,...,s,    
\ee
are in $\cal R$. It follows from Lemma 6.2 in \cite{sussJ}
that every matrix in $e^{\cal L}$ can be expressed 
as a finite product of  matrices of
the form $e^{At}$, with $A \in \{ A_1,...,A_s\}$ and $t \in
\RR$. Moreover, given a matrix of the form $e^{A t}$, with $t \in \RR$,
by compactness of the Lie group $U(n)$ (see e.g. the argument in Theorem
6.5 of \cite{sussJ}) we can always choose a {\it positive} $\alpha$ such that
$e^{A \alpha}$ is arbitrarily close to $e^{At}$. Combining these 
arguments it follows that $\cal R$ is dense in $e^{\cal L}$. 

\vs 

{\it Fact 3:} { $\cal R$ contains a neighborhood of the 
Identity in $e^{\cal L}$}

\vs 
 
Let $n$ be the dimension of $\cal L$. We can choose
(see Lemma 1 in \cite{miko}) $n-s$ elements in $e^{\cal L}$,
$U_1,...,U_{n-s}$  and $n-s$ element in the set $\{A_1,...,A_s\}$, say
$\bar A_1,...,\bar A_{n-s}$,  such that,  
\be{base}
\{ A_1,A_2,....,A_s, U_1\bar A_1 U_1^{-1},...., U_{n-s} \bar A_{n-s}
U_{n-s}^{-1} \}, 
\ee 
form a basis of $\cal L$.
Given n elements $V_1,....,V_n$ in ${\cal R}$,  consider the
function $\Phi(j_1,...,j_n)$ from a neighborhood of the point
$(1,...,1)$  in $\RR^n$ to $e^{\cal L}$,     
\be{funzione}
\Phi(j_1,...,j_n):=e^{A_1j_1} V_1 e^{A_2j_2} V_2 \cdot \cdot \cdot
e^{A_sj_s} V_s e^{\bar A_1j_{s+1}}
V_{s+1} \cdot \cdot \cdot e^{\bar A_{n-s}j_n} V_{n}. 
\ee
All the elements obtained by $\Phi(j_1,...,j_n)$ with $j_1,...,j_n \geq
0$ are in ${\cal R}$. We can show that, if we choose appropriately
$V_1,...,V_n$,  these elements contain a
neighborhood of the point $U_0:=e^{A_1} V_1 e^{A_2} V_2 \cdot \cdot \cdot
e^{A_s} V_s e^{\bar A_1}
V_{s+1} \cdot \cdot \cdot e^{\bar A_{n-s}} V_{n} $ by showing
that the Jacobian at $(1,...,1)$ is non zero and applying the implicit
functions theorem (cfr. \cite{Weaver}).  
\vs

Differentiating at $(1,...,1)$ the function 
$\Phi(j_1,...,j_n)$, we obtain

\begin{eqnarray}
\frac{\partial \Phi(j_1,...,j_n)}{\partial j_1}|_{j_1,...,j_n=1,...,1}
=A_1 e^{A_1} V_1 e^{A_2}V_2 \cdot \cdot \cdot e^{A_s}V_s e^{\bar A_1}
V_{s+1} \cdot \cdot \cdot e^{\bar A_{n-s}}V_n,  \\
\frac{\partial \Phi(j_1,...,j_n)}{\partial j_2}|_{j_1,...,j_n=1,...,1}
=e^{A_1} V_1 A_2 e^{A_2}V_2 \cdot \cdot \cdot e^{A_s}V_s e^{\bar A_1}
V_{s+1} \cdot \cdot \cdot e^{\bar A_{n-s}}V_n, \nonumber\\
\cdot \nonumber \\
\cdot \nonumber \\
\cdot \nonumber\\
\frac{\partial \Phi(j_1,...,j_n)}{\partial j_{s+1}}|_{j_1,...,j_n=1,...,1}
=e^{A_1} V_1 e^{A_2}V_2 \cdot \cdot \cdot e^{A_s}V_s \bar A_1 e^{\bar A_1}
V_{s+1} \cdot \cdot \cdot e^{\bar A_{n-s}}V_n, \nonumber \\
\cdot  \nonumber\\
\cdot  \nonumber\\
\cdot  \nonumber \\
\frac{\partial \Phi(j_1,...,j_n)}{\partial j_{s+1}}|_{j_1,...,j_n=1,...,1}
=e^{A_1} V_1 e^{A_2}V_2 \cdot \cdot \cdot e^{A_s}V_s  e^{\bar A_1}
V_{s+1} \cdot \cdot \cdot \bar A_{n-s} e^{\bar A_{n-s}}V_n.  \nonumber
\end{eqnarray}

Choosing 
\begin{eqnarray}
V_1 \approx e^{-A_1}, \\
\cdot \nonumber \\
\cdot \nonumber \\
\cdot \nonumber \\
V_{s-1} \approx e^{-A_{s-1}}, \nonumber \\
V_s \approx e^{-A_{s}} U_1, \nonumber \\
V_{s+1} \approx e^{- \bar A_1} U_1^{-1} U_2, \nonumber \\
\cdot \nonumber \\
\cdot \nonumber \\
\cdot \nonumber \\
V_{n-1} \approx e^{- \bar A_{n-s-1}} U_{n-s-1}^{-1} U_{n-s}, \nonumber \\
V_n \approx e^{- \bar A_{n-s}} U^{-1}_{n-s},  \nonumber 
\end{eqnarray}
we obtain 
\begin{eqnarray}
\frac{\partial \Phi(j_1,...,j_n)}{\partial j_{1}}|_{j_1,...,j_n=1,...,1}
\approx A_1, \\
\cdot \nonumber \\
\cdot \nonumber \\
\cdot \nonumber \\
\frac{\partial \Phi(j_1,...,j_n)}{\partial j_{s}}|_{j_1,...,j_n=1,...,1}
\approx A_s,  \nonumber \\
\frac{\partial \Phi(j_1,...,j_n)}{\partial j_{s+1}}|_{j_1,...,j_n=1,...,1}
\approx U_1 \bar A_1 U_1^{-1},  \nonumber \\
\cdot \nonumber \\
\cdot \nonumber \\
\cdot \nonumber \\
\frac{\partial \Phi(j_1,...,j_n)}{\partial j_{n}}|_{j_1,...,j_n=1,...,1}
\approx U_{n-s} \bar A_{n-s} U_{n-s}^{-1},  \nonumber \\
\end{eqnarray} 
which are linearly independent and this proves our claim.

\vs 
 
Now, from the fact that the open set $B(U_0, \epsilon)$ is 
a subset of $ {\cal R}$ it follows that $\cal R$ contains a neighborhood of
the Identity. In particular choose $U_0^{\bar k} \in {\cal R}$,  such that
 $$ || U_0^{\bar k}- U_0^{-1} || < \frac{\epsilon}{2}.$$ Then if $F \in
B(I, \frac{\epsilon}{2}) $, writing $F=U_0^{\bar k} X$ we have 
\be{inequalities}
\frac{\epsilon}{2} > ||F- I|| \geq || X-U_0|| - ||U_0^{\bar k} -
U_0^{-1}||, 
\ee
and therefore 
\be{sf}
|| X-U_0|| < \epsilon, 
\ee
which implies $X \in \cal R$ end therefore $F=U_0^{\bar k} X \in {\cal
R}$ as well. 

\vs 

{\it  Conclusion:}

\vs 

The semigroup $\cal R$ contains a neighborhood of the identity in
$e^{\cal L}$. Since $e^{\cal L}$ is connected, it follows that 
${\cal R}=e^{\cal L}$. \qed

\vs

Notice we only used compactness of the Lie group $U(n)$ and therefore
the proof will go through if $e^{\cal L}$ was a connected Lie subgroup
of a general compact Lie Group.

\end{document}